\documentclass[aps,prb,amsmath,amssymb,twocolumn,superscriptaddress,showpacs,floatfix]{revtex4-1}
\usepackage{ifpdf}
 \newif\ifpdf
\ifx\pdfoutput\undefined
   \pdffalse
\else
   \pdfoutput=1
   \pdftrue
\fi
\ifpdf
   \usepackage{graphicx}
   \usepackage{epstopdf}
\else
   \usepackage{graphicx}
\fi

\usepackage{bm}
\usepackage{epsfig}
\usepackage[usenames]{color}
\usepackage{soul}
\usepackage{graphicx}
\usepackage{hyperref}
\newcommand{\Cs}{\mathit C_{\scriptscriptstyle{\Sigma}}}

\DeclareMathOperator{\sign}{sign}
\DeclareMathOperator{\Real}{Re}

\begin{document}

\title{Memory effect in ferroelectric single electron transistor: violation of conductance periodicity in the gate voltage}

\author{S.~A.~Fedorov}
 \affiliation{Department of Theoretical Physics, Moscow Institute of Physics and Technology, Moscow 141700, Russia}
 \affiliation{P.N. Lebedev Physical Institute of the Russian Academy of Sciences, Moscow 119991, Russia}
\author{A.~E.~Korolkov}
\affiliation{Department of Theoretical Physics, Moscow Institute of Physics and Technology, Moscow 141700, Russia}
\affiliation{P.N. Lebedev Physical Institute of the Russian Academy of Sciences, Moscow 119991, Russia}
\author{N.~M.~Chtchelkatchev}
\affiliation{Department of Theoretical Physics, Moscow Institute of Physics and Technology, Moscow 141700, Russia}
\affiliation{Department of Physics and Astronomy, California State University Northridge, Northridge, CA 91330, USA}
\author{O.~G.~Udalov}
\affiliation{Department of Physics and Astronomy, California State University Northridge, Northridge, CA 91330, USA}
\affiliation{Institute for Physics of Microstructures, Russian Academy of Science, Nizhny Novgorod, 603950, Russia}
\author{I.~S.~Beloborodov}
\affiliation{Department of Physics and Astronomy, California State University Northridge, Northridge, CA 91330, USA}

\date{\today}

\begin{abstract}
The fundamental property of most single-electron devices with quasicontinuous quasiparticle spectrum on the island is the periodicity of their transport characteristics in
the gate voltage. This property is robust even with respect to placing the ferroelectric insulators in the source and drain tunnel junctions. We show that placing the
ferroelectric inside the gate capacitance breaks this periodicity. The current-voltage characteristics of this SET strongly depends on the ferroelectric polarization and shows the giant memory-effect even for negligible ferroelectric hysteresis making this device promising for memory applications.
\end{abstract}

\maketitle

\section{Introduction}

Ferroelectricity like magnetism has been under investigation for a decades. Recent progress in ferroelectricity is stimulated
by i) miniaturisation of ferroelectric samples to nanoscale where they show new physical properties compared to the bulk ferroelectric materials,~\cite{dawber2003self,ahn2004ferroelectricity,Dawber2005RevModPhys,wang2007modeling,zhang2007improved,Scott2007,Maksymovych2009Science,chu2008Nature,lee2008nanocapacitor,kalinin2010RPP,Dawber2012,Dawber2012_1,ortega2012relaxor,Chanthbouala2012} and ii) modern computer processors and memory demand storing and moving electric charges, and control associated electric fields. There is a tendency for further increase of computer efficiency that results in facing the nanolevel with individual electrons and atoms. In ferroelectric materials polarization is produced by atom displacements. Nanoferroelectrics combined with single electron-nanocircuits are thus promising devices for memory storage and information processing.

Recently it was shown that the presence of ferroelectricity in the source and drain tunnel junctions of single electron transistor (SET) induces memory effect in the current-voltage characteristics even in the limit of negligible hysteresis of the ferroelectric insulators.~\cite{RefOurPRB} The bottle-neck of ferroelectric SET is the experimental difficulty to produce ultrathin and ultra small ferroelectric tunnel junctions with special parameters. Technologically it is much easier to produce a nanothin ferroelectric layer but not thin enough for perfect electron tunneling. Such layer can be placed into the gate capacitance of the SET. From the first glance physics of such ferroelectric should be similar to physics of SET considered in Ref.~\onlinecite{RefOurPRB} with ferroelectric source and drain capacitors. But this is not so.

The fundamental property of most single-electron devices with quasicontinuous quasiparticle spectrum on the island is the periodicity of their transport characteristics in
the gate voltage. This property is robust even with respect to placing the ferroelectric insulators in the source and drain tunnel junctions.~\cite{RefOurPRB} We show that placing the ferroelectric inside the gate capacitance breaks this periodicity even for negligible ferroelectric hysteresis. Applying relatively small ``switching'' gate voltage one can change the polarization of ferroelectric. We show that further increase of the gate voltage does not affect the direction of ferroelectric polarization. The current-voltage characteristics of this SET strongly depends on the ferroelectric polarization and shows the memory-effect making this device promising for memory applications.
\begin{figure}[bt]
  \centering
  \includegraphics[width=0.98\columnwidth]{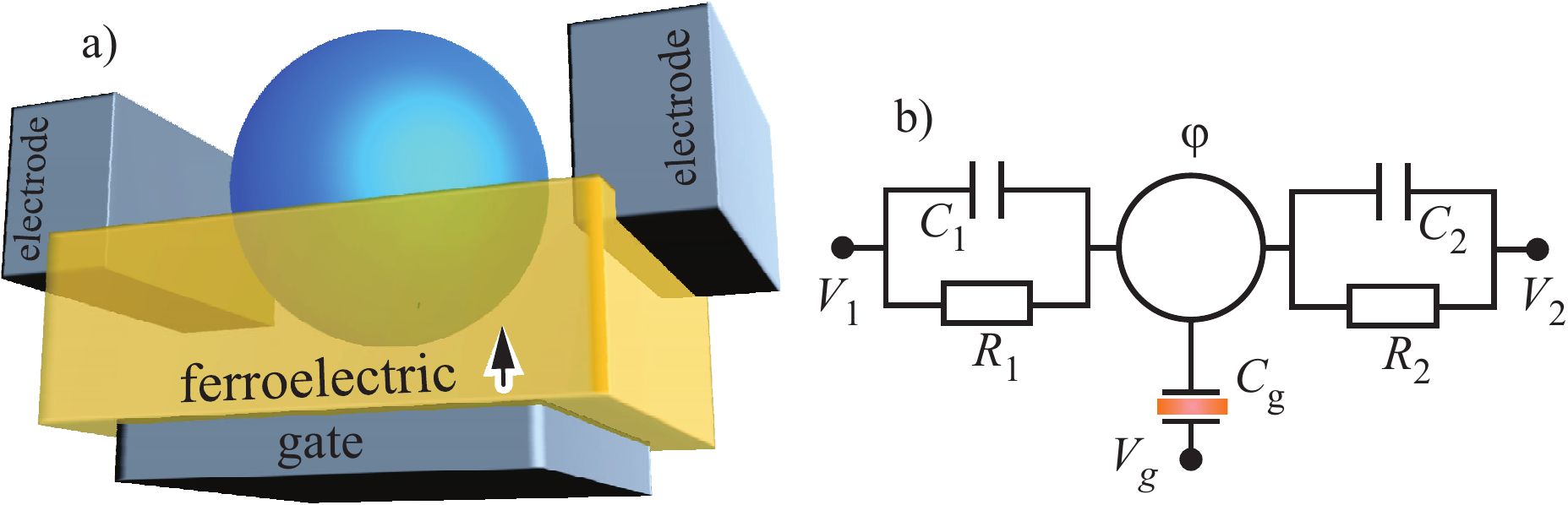}\\
  \caption{(Color online) (a) Sketch of single electron transistor (SET)
  with ferroelectric placed at the gate capacitor. (b) The equivalent scheme.}\label{fig_device}
\end{figure}

\section{Single electron transistor with ferroelectric gate}

\subsection{The Model}

The ferroelectricity localized in the gate-capacitance changes the distribution of the excess
charge in the nano-grain. In the absence of quantum fluctuations the grain charge is $ne = \sum_i\left \{C_i\left[\phi(n)-V_i\right]\right \}+\int_g d\mathbf S_g\cdot \mathbf P_g$. Here  $n$ is excess charge number, $e$ is the elementary charge, $\phi(n)$ is the potential of the nano-grain, $C_i$ with $i=1,2,g$ are the capacitances and $\mathbf S_g$ is the grain surface. The surface integration is performed over the nano-grain part which is in contact with the ferroelectric. The polarization $\mathbf P_g$ itself depends on the grain charge. Thus, calculations of the charge statistics and
polarization should be done self-consistently.~\cite{RefOurPRB}

\begin{figure}[tb]
  \centering
  \includegraphics[width=0.8\columnwidth]{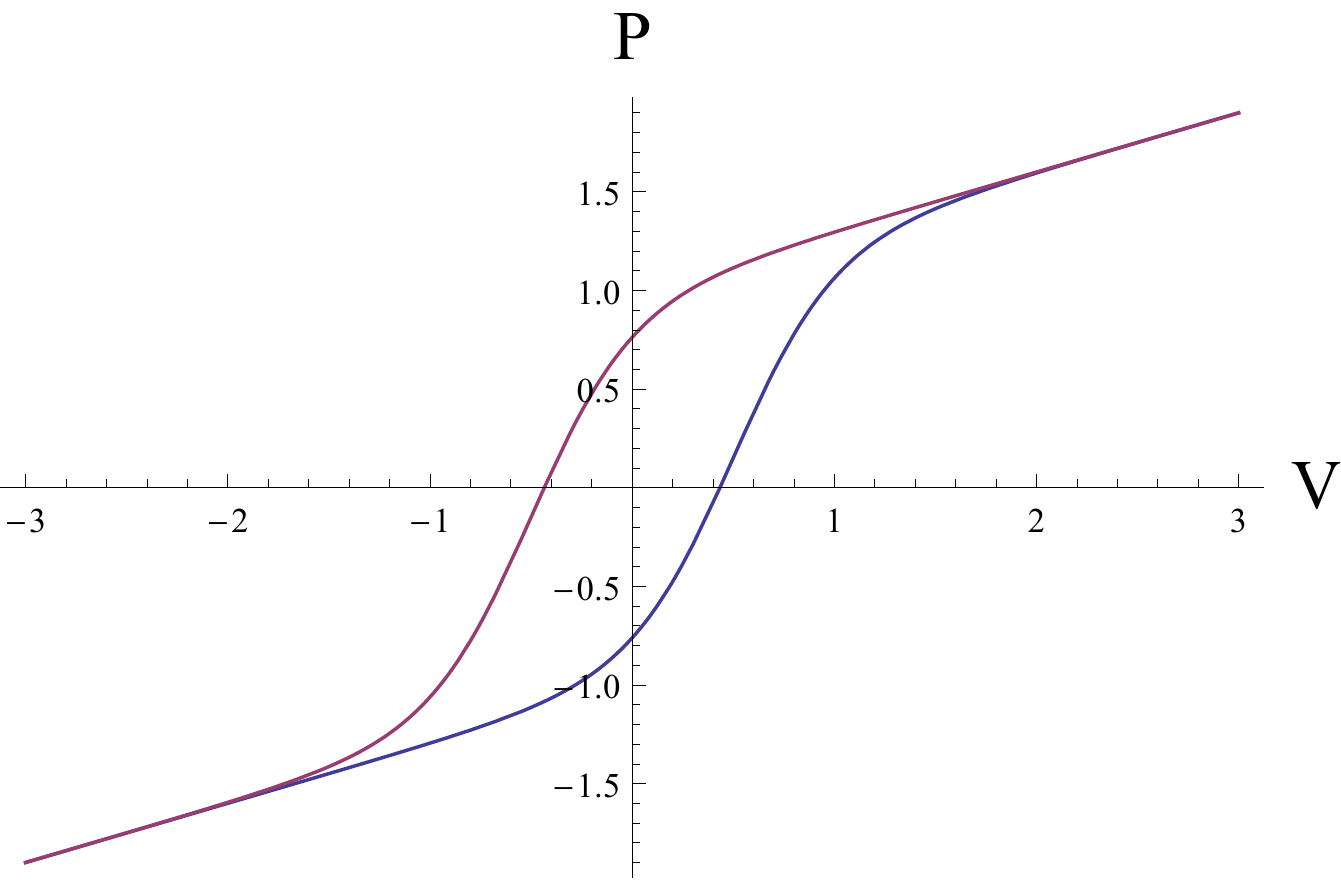}\\
  \caption{(Color online) Model polarization}\label{fig_PE}
\end{figure}

Generally the electric field dependence of polarization in SET has hysteresis.
The following model takes this effect into account
\begin{gather}\label{P-approx}
P^{(u/d)}(\mathcal E)= P^{0}\tanh\left(\frac{\mathcal E\pm \mathcal E_h}{\mathcal E_{s}}\right)+\alpha \mathcal E,
\end{gather}
where ``u'' and ``d'' stands for the upper and lower branches of hysteresis loop, $\mathcal E_{s}$ is the saturation field, $P^{0}$ is the saturation polarization amplitude, $\mathcal{E}_h$ describes the width of the hysteresis loop. Similarly we can write the voltage dependence of the polarization introducing $V_s=\mathcal E_s d$ and $V_h=\mathcal E_h d$ where $d$ is the
width of the gate capacitor: $P^{(u/d)}(V)= P^{0}\tanh\left(\frac{V\pm V_h}{V_{s}}\right)+\alpha V$.
The typical graph of $P(V)$ is shown in Fig.~\ref{fig_PE}.

In typical single-electron devices the electron tunneling is a much faster
process then the relaxation of the ferroelectric.~\cite{RefOurPRB} This is related to the fact that
ferroelectricity is related to the shift of ions that are heavy and inert. Therefore the ferroelectric in SET is sensitive to the average electric field over the ``fast'' electron-tunneling events.~\cite{RefOurPRB} Thus, for a fixed ferroelectric polarization $\mathbf P$ we can calculate the average grain potential $\langle\phi\rangle$. Since $\langle\phi\rangle$ itself depends on the polarization $P$ through the probability distribution $p(n)$ to find $n$ excess charges on the grain we obtain the self-consistency equation. The probabilities itself can be calculated
using the modified SET Orthodox theory, see Ref.~\onlinecite{RefOurPRB} and Refs.~\onlinecite{Armour2004PRB,Nishiguchi2008PRB,Korotkov1994PRB} where selfconsistency was developed
for SET with slowly oscillating gate electrode.
\begin{figure*}[t]
  \centering
  \includegraphics[width=\textwidth]{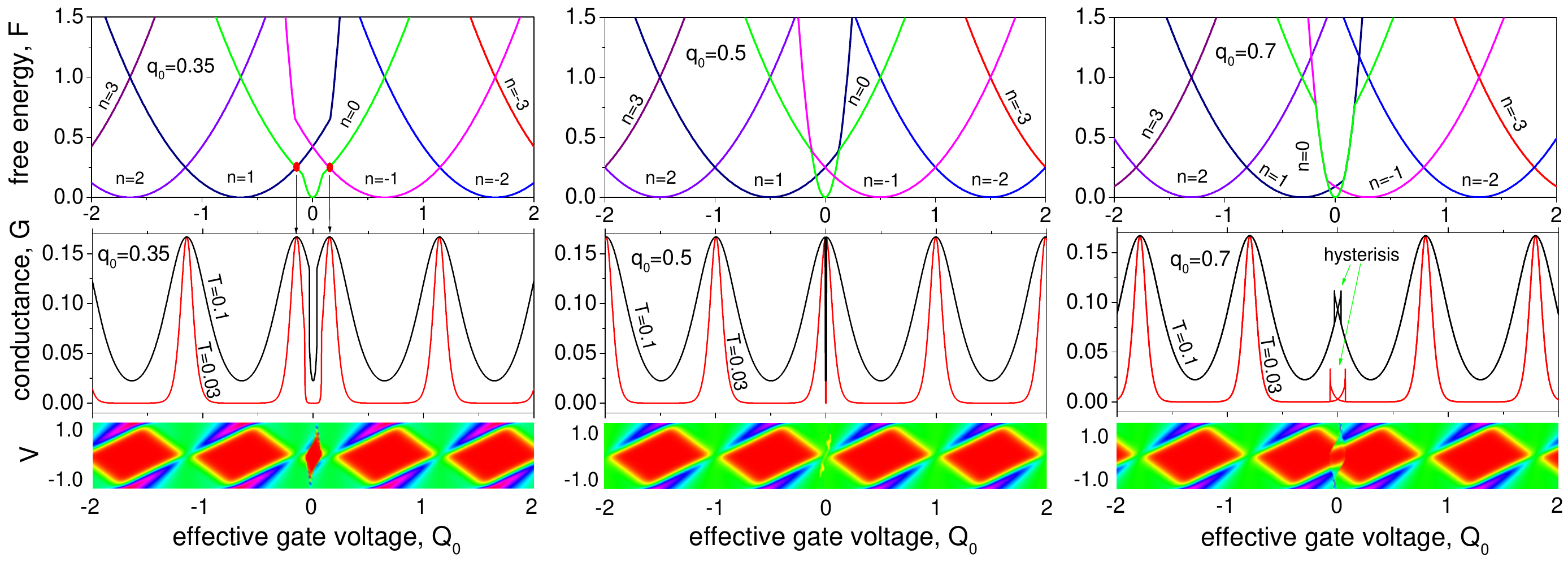}\\
  \caption{(Color online) Ferroelectricity breaks down the fundamental property of SET: periodicity of its transport properties in the gate voltage $V_g$.
  Only for effective gate voltage $|Q_0|>1/2$ the conductance peaks are equidistant. Upper row: the effective free energy $f_n$ of SET, see Eq.~\eqref{eqFparabola}. Middle raw: the zero bias conductance. Lower raw: the conductance density plot. Parameters: $C_1=0.3$, $C_2=0.5$, $C_g=0.2$, $R_1=1$, $R_2=2$, $V_s=0.01$, and $V_h=0$.  }\label{figenergy}
\end{figure*}

\subsection{Dimensionless units and basic notations\label{secUnits}}

SET has a number of important parameters. It is convenient to use the
following units for analytical and numerical calculations: $2E_c=e^2/\Cs$ is the energy and temperature unit ($k_B=1$), the elementary charge $e$ is the charge-unit (electron charge  is equal $-1$). The capacitance becomes dimensionless using $e^2/2E_c$, so $\Cs=1$. We use the (bare) tunnelling resistance of the first tunnel junction, $R_1$, between the left electrode and the nanograin as the unit for the resistance. Also we use the effective gate charge
as a control parameter, $Q_0=-V_g C_g$.

\section{Ferroelectric SET at low temperatures. Orthodox theory.}

Here we consider the limit of low temperatures, $T \ll E_c$ where
transport properties of SET
(not far from the degeneracy points) are well described by the ``orthodox model''.~\cite{averin1991single,averin1991theory,devoret1992single,wasshuber2001computational} Generalization of the orthodox model for SET with ferroelectricity was recently formulated in Ref.~\onlinecite{RefOurPRB}. Using equations of Ref.~\onlinecite{RefOurPRB} one can find the distribution functions $p(n)$ describing the excess charge statistics on the grain, the mutual dependence of ferroelectric polarization, the charge statistics and electron current.\par

\subsection{Ferroelectric with ultrathin hysteresis loop}

First we consider the limiting case of ultrathin ferroelectric hysteresis loop,
with $\mathcal E_h=0$ in Eq.~\eqref{P-approx}. Also we assume that the parameter $\alpha=0$
in Eq.~\eqref{P-approx}. Using these assumptions we explain the strong influence of ferroelectricity
in the gate capacitor on the fundamental property of SET: periodicity of its transport properties
in the gate voltage $V_g$.

\subsubsection{Ferroelectric with small switching field: $V_s\ll 1$}

The basic features of SET with ferroelectricity in the gate capacitor can be qualitatively
understood for small switching field, $V_s\ll 1$.

First we investigate the SET for zero driving voltage, $V=0$, and zero switching voltage, $V_s=0$. We also assume that electrons do not hop between leads and the grain. This assumption is reasonable for gate voltage being away from the points of intersection of different free energy branches. In this case the charge number on the grain does
not fluctuate and the average potential is equal to the
instant potential $\phi=\langle\phi\rangle$ and $\int_g \textbf P \cdot d\textbf S_g=q_0\sign(\phi-V_g)$. The equation describing the potential $\phi$ has the form
\begin{equation}\label{EqCharge0}
ne=C\phi+C_g(\phi-V_g)+q_0\sign(\phi-V_g).
\end{equation}
Here $C=C_1+C_2$. Equation~\ref{EqCharge0} has the following solution in the limit $V_s\to 0$:
\begin{equation}\label{EqCharge}
\phi=\left\{\begin{array}{l}
\frac{C_gV_g+ne+q_0}{C+C_g}, \textrm{~} V_g>\frac{ne+q_0}{C},\\
\frac{C_gV_g+ne-q_0}{C+C_g}, \textrm{~} V_g<\frac{ne-q_0}{C},\\
V_g, \textrm{~} \frac{ne-q_0}{C}<V_g<\frac{ne+q_0}{C}.
\end{array}\right.
\end{equation}
Each branch of potential $\phi$ has three regions. Two regions have the same slope but their shifted by the voltage
$2q_0/\Cs$. The transition between these regions is located in the vicinity of the point $Q_0=-C_g ne/C$ with
the width of $C_g 2q_0/C$. For parameter $q_0<0.5e$ the different branches do not intersect while for $q_0>0.5e$ the
branches intersect on the line $\phi=-Q_0/C_g$. The intersection of $\phi$ with $0$ is happening at
the point $(ne\pm q_0)$.

Following the orthodox theory we introduce the effective free energy of the SET
\begin{gather}\label{eqFparabola}
  F=\min_n f_n, \,\, f_n=E_c(C_\Sigma\phi)^2=E_c(e n-Q_0-P)^2.
\end{gather}

For zero polarization, $P = 0$, the function $f_n$ is a parabola in parameter
$Q_0=-V_g C_g$. At finite $P$ the ferroelectric polarization depends on the parameter $Q_0$
making the function $f_n(Q_0)$ more complicated function consisting of parabola
fragments separated by some transition region, see Fig.~\ref{figenergy}. At the degeneracy
points, where the function $f_n(Q_0)$ intersects, the Coulomb blockade is suppressed
allowing electrons go through the SET from one lead to another. The positions of these degeneracy
points, corresponding to the conductivity maximums, is $((n+1/2)e\pm q_0)$. Thus, all the
conductivity peaks move to the zero gate voltage point. If a peak reaches the point $Q_0=0$
it stays at this point with further increase of parameter $q_0$.
All peaks have the same shift magnitude, but different direction for peaks below and above
the point $Q_0=0$. This leads to the breaking of periodicity in the conductivity peaks of FE SET.  The numerically calculated conductance peaks approximately correspond to the degeneracy points as shown in Fig.~\ref{figenergy}.

In orthodox theory the conductance of SET (without ferroelectricity) at low temperatures, $T\ll E_c$, is
\begin{gather}\label{go}
G(\delta Q_0)=\frac 12 \cdot \frac 1{R_1+R_2} \cdot \frac{e\,\delta Q_0/\Cs T}{\sinh(e\,\delta Q_0/\Cs T)}.
\end{gather}
Here $\delta Q_0 = \min_k[Q_0 - (2k+1)\frac e2] \ll e$ is the deviation of the induced charge by the gate terminal from the nearest degeneracy point. With ferroelectricity in the gate, we should replace the deviation $\delta Q_0$ by
\begin{multline}\label{Qs1}
\delta Q_s \to \min_k\left[q_0\tanh\left(\frac{\langle\phi\rangle+Q_0/C_g}{V_s}\right)+\right.
\\
\left.Q_0- (2k+1)\frac e2\right].
\end{multline}
This equation is valid for small parameter $q_0 \ll 1$. For small switching voltage, $V_s\lesssim1$, we can replace
$\tanh$ in Eq.~\eqref{Qs1} by unity and using
Eq.~(\ref{go}) find that the conductance peaks are shifted by $q_0$ for $Q_0 > 0$ and by $-q_0$ for $Q_0<0$.
This is consistent with numerical calculations, see Fig.~\ref{figenergy}.

The most important effect that follows from Eq.~\eqref{Qs1} in the presence of ferroelectric
is the break up of conductance periodicity
in the parameter $Q_0$. This periodicity is the basic property of SETs;
it is robust for ferroelectrics being present in the capacitors
between the left and the right leads.~\cite{RefOurPRB} However, numerical calculations
show that this periodicity is absent for ferroelectric
being place inside the gate capacitor, see Fig.~\ref{figenergy}.\par

This break of periodicity is a very general result. It follows
from the non-periodic and non-linear dependence of the FE polarization on the
effective charge $Q_0$. The FE polarization is defined by
the difference of two quantities, the average grain potential $\langle\phi\rangle$ and the gate voltage $V_g$.
The potential $\langle\phi\rangle$ oscillates around zero while the voltage $V_g$ grows unlimited. As a result, the FE polarization is saturated for voltages $|V_g|\gg|e|/C_{\Sigma}+V_s$, however
its direction depends on the sign of the gate voltage $V_g$ producing the opposite shifts of
conductance peaks at voltages $V_g=\pm\infty$.\par

\begin{figure}[t]
  \centering
  \includegraphics[width=\columnwidth]{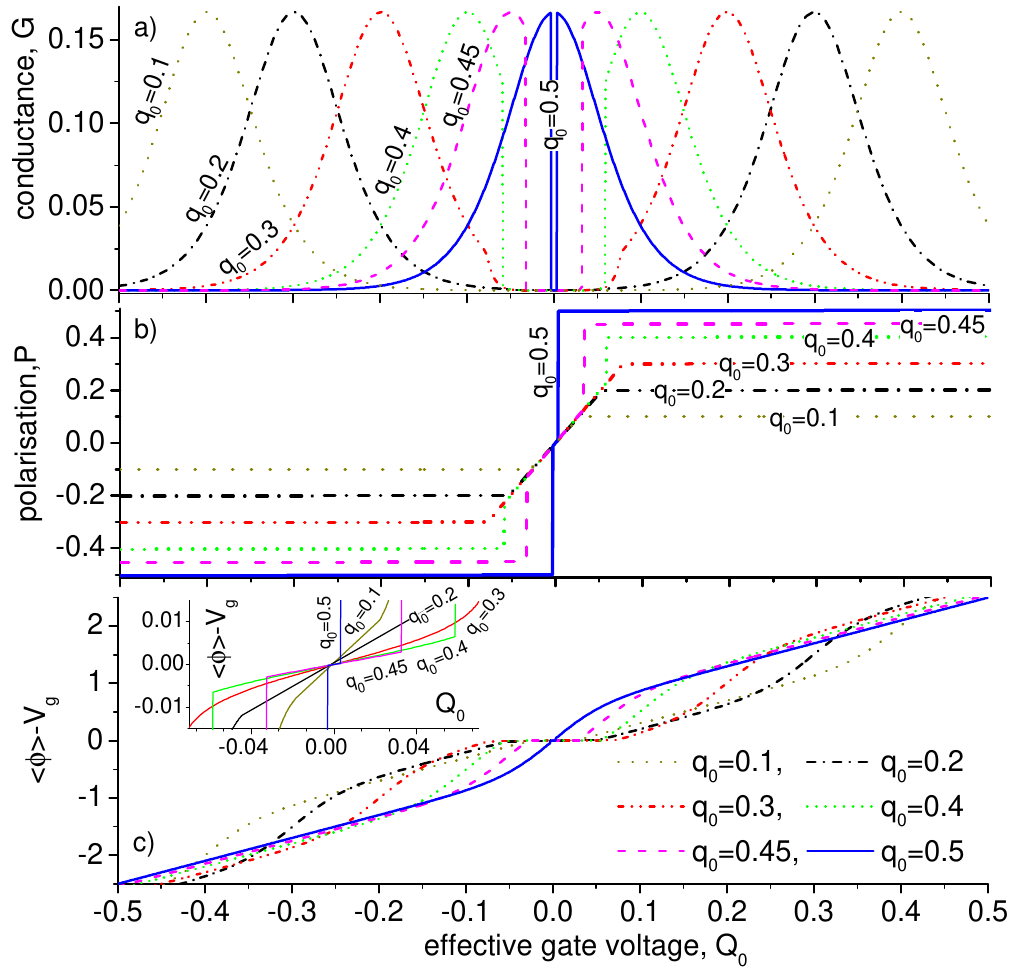}\\
  \caption{(Color online) a) Conductance vs effective gate voltage $Q_0$. b) Ferroelectric polarization vs $Q_0$.
  c) Voltage bias in the gate capacitor, $\langle\phi\rangle-V_g$. The jumps in the conductance and polarization in Fig.~\ref{figenergy} for $q_0<0.5$ are related to the memory effect instability but with
  thin hardly detectable hysteresis loop. Insert:  voltage bias vs $Q_0$ for
  small values of $Q_0$. All parameters are the same as in Fig.~\ref{figenergy}, except $T=0.03$. }\label{figGP}
\end{figure}
\begin{figure}[t]
  \centering
  \includegraphics[width=\columnwidth]{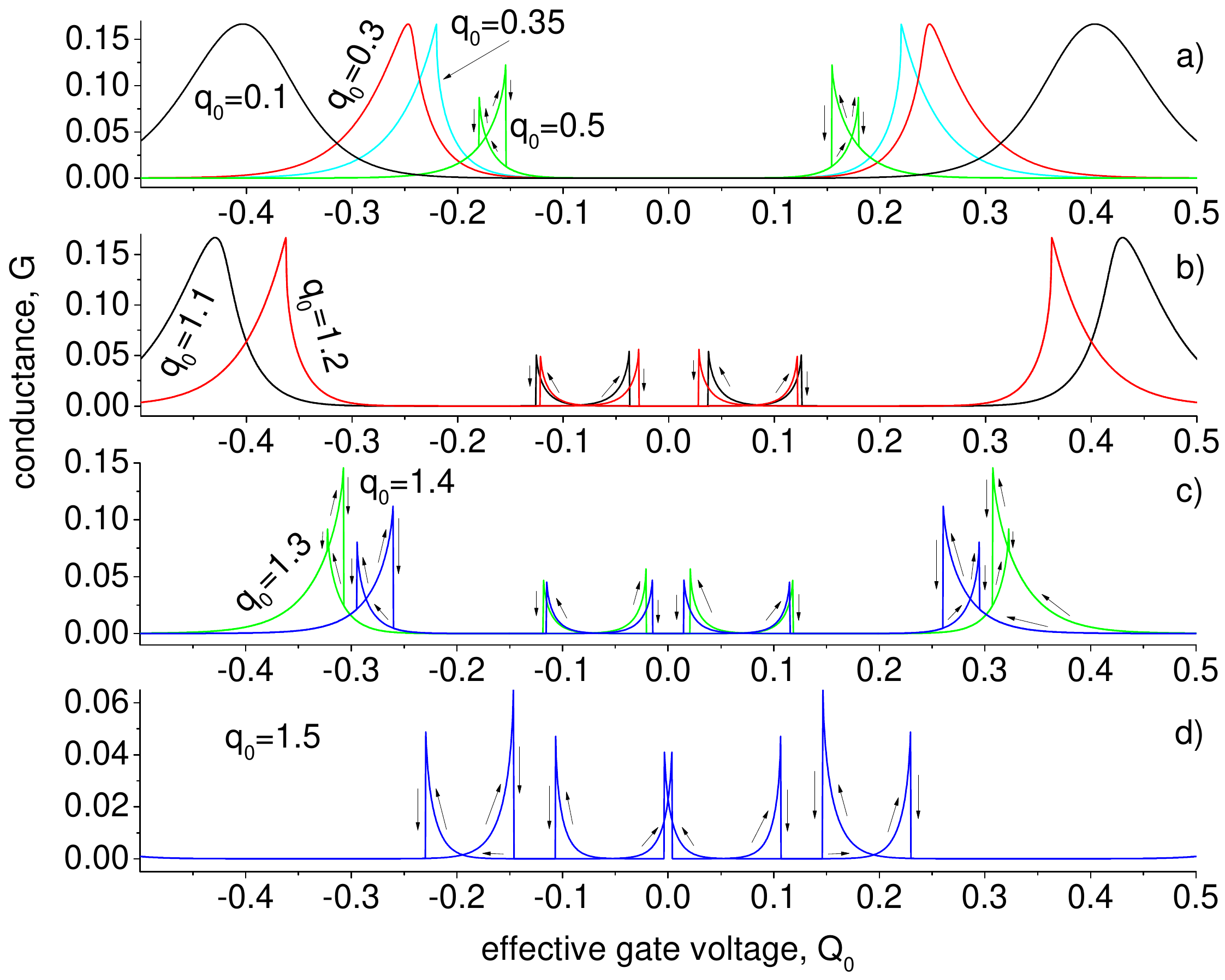}\\ \includegraphics[width=0.95\columnwidth]{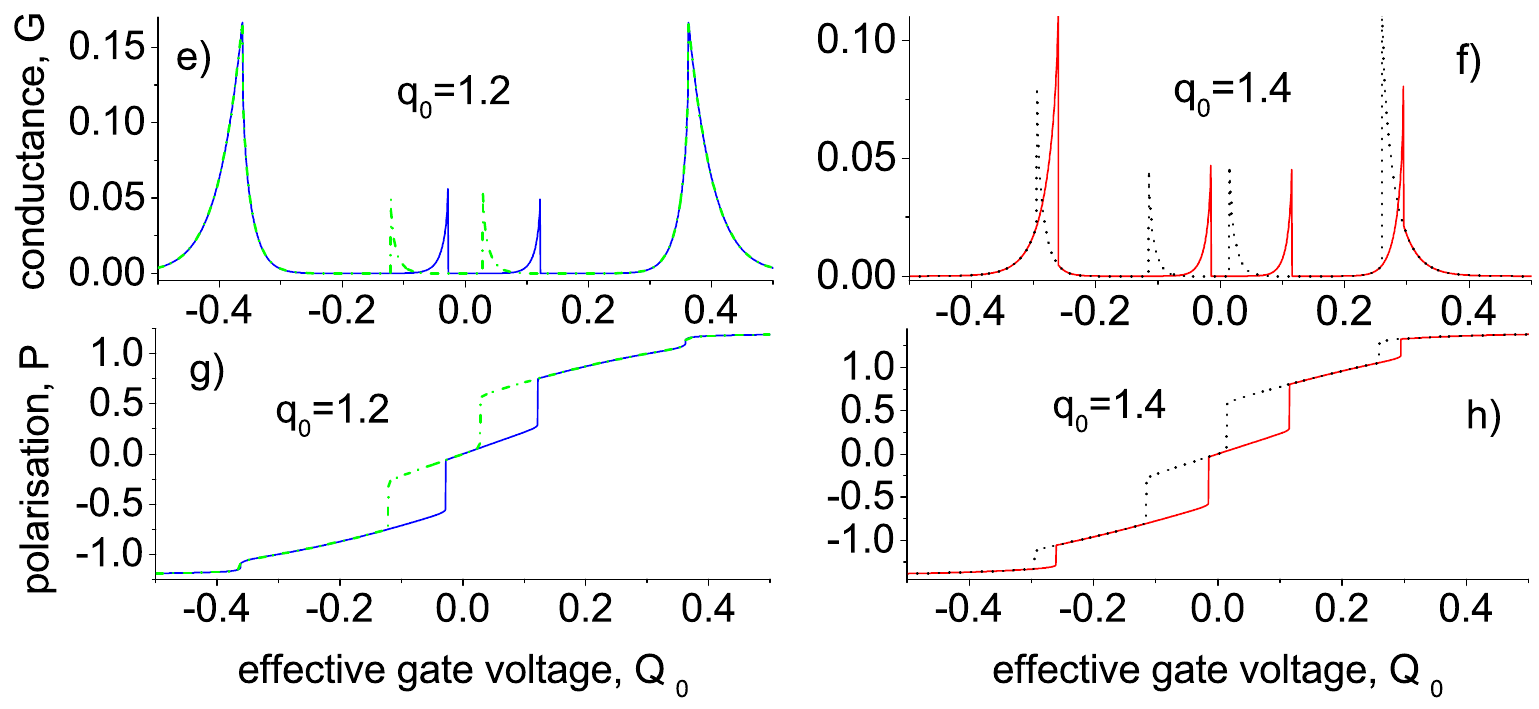}\\
  \caption{(Color online) Conductance $G$ and polarisation $P$ vs $q_0$ for $V_s=1$. All parameters are the same
  as in Fig.~\ref{figenergy}, except $T=0.03$. With increasing $q_0$ the conductance peaks
  are shifting towards $Q_0=0$. At critical $q_0$ the bifurcation appears and conductance peaks
  close to $Q_0=0$ become memory-dependent. (a) For $q_0<0.35$ the conductance $G(Q_0)$ is
  ambiguous function. At $q_0\approx0.35$ the bifurcation point for the first
  pair of $G$-peaks appears. (b) At $q_0\approx 1.2$ the second bifurcation
  point for the next pair of conductance peaks appears.
  (c) Evolution of the memory effect with $q_0$.
  (d) For further increase of $q_0$ the memory-dependent conductance peaks are shifting
  towards $Q_0=0$ and finally ``collide'' with each other. Arrows in (a)-(d) show
  the memory effect branches for increasing (decreasing) $Q_0$.  (e-h) Conductance
  and polarization of the ferroelectric. The branches of $G(Q_0)$ for increasing
  (decreasing) $Q_0$ are shown by solid and discontinuous curves, respectively.}\label{figVs1G}
\end{figure}

Another interesting phenomenon appearing due to the presence of the
FE layer is the hysteresis conductivity behavior. We remind that in this section we consider the
FE without hysteresis. However, even in this case the conductivity peak in the vicinity of $Q_0=0$
is split into two branches, see the right panel in Fig.~\ref{figenergy}. The
pronounced~\cite{footnote} hysteresis appears for $q_0>0.5$ where the system polarization has two stable ground states corresponding
to two different directions of the FE polarization (toward the grain and toward the gate) and two
different grain charges, positive and negative. These two states have different conductivity.
For step-like polarization the hysteresis appears for peaks located in the vicinity of $Q_0=0$.

With increasing temperature the hysteresis disappears.
The criterion for hysteresis existence has the form,
$e\Gamma_{0,1}/(\Gamma_{0,1}+\Gamma_{1,0})=Q_{\mathrm{max}}<q_0$, where $\Gamma_{i,j}$ is the transition rate between the states with grain charges $i$ and $j$ and $0.5<q_0<1.5$. $Q_{\mathrm{max}}$ decreases with increasing temperature.
We demonstrate the existence of two ground states in the Appendix~\ref{sec:App_Hyst1}.

\textit{Now we consider small but finite values of switching voltage $V_s$} and investigate
the correlation of the conductance peaks and polarization evolution with parameter $Q_0$. For
parameter $q_0\leq 0.5$ the conductance peaks do not merge, however they deform
approaching the point $Q_0=0$. This is related to the switching of polarisation with $Q_0$, see Fig.~\ref{figGP}.

For $0.5<q_0<1$ the conductance shows the pronounced memory effect (hysteresis), Fig.~\ref{figenergy}. The positions of the jumps in the conductance correspond to the divergences of $\frac{dG}{dQ_0}$.~\cite{RefOurPRB} The amplitude of the conductance peaks in this regime is suppressed for $q_0 > T$. The evolution of memory effect in the
conductance with parameter $q_0$ is shown in Fig.~\ref{figenergy}. The conductance has similar behavior
for $1<q_0<1.5$ and for $1.5<q_0<2$ its behavior coincide with conductance behavior
for $0.5<q_0<1$. Thus, the transport properties of SET are periodic in $q_0$ with the
period $1$. However, this statement is approximate:
it is valid for small values of switching voltage $V_s$ and a negligible parameter $\alpha$ only.

\begin{figure*}[t]
  \centering
  \includegraphics[width=2.07\columnwidth]{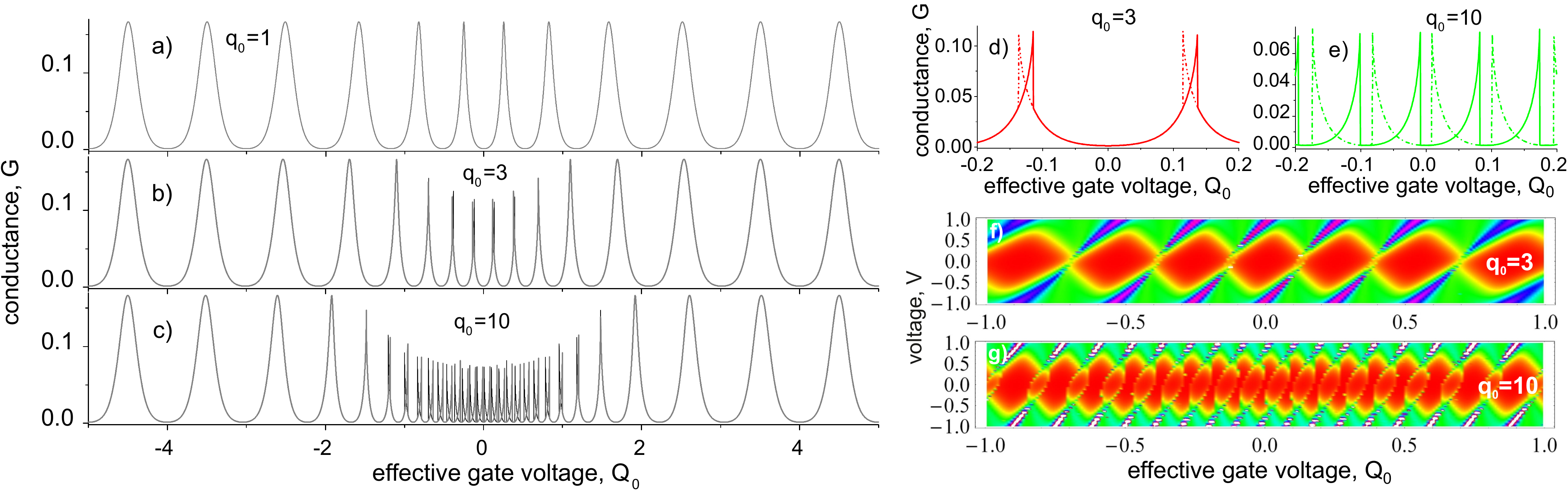}\\
  \caption{(Color online) SET device at saturation voltage $V_s=5$. Parameters are the same
  as in Fig.~\ref{figenergy}. (a)-(c) Conductance evolution with $q_0=1,3,10$. (d) and
  (e) Fine structure of the conductance peaks for $q_0=3,10$.
  Branches of $G(Q_0)$ for increasing (decreasing) $Q_0$ are shown
  by solid and discontinuous curves. (f) and (g) Conductance density plots;
  the Coulomb island-structures strongly differs from the SET without ferroelectricity.}\label{DplotVs10}
\end{figure*}

The insert in Fig.~\ref{figGP} shows the deviation of $\langle \phi\rangle$ from $V_g$ in the region, $(ne-q_0)/C\lesssim V_g\lesssim (ne+q_0)/C$. It nearly linearly depends on $Q_0$.  After the substitution of $\tanh\left(\frac{\langle\phi\rangle-V_g}{V_{s}}\right)$ instead of $\sign(\phi-V_g)$ in Eq.~\eqref{EqCharge0} the estimate follows that in the leading order over $V_s$, $\langle \phi\rangle-V_g\approx V_s Q_0/q_0$.

\subsubsection{Ferroelectric with large switching field: $V_s\gtrsim 1$.
Breaking and condensation of conductance peaks.}

For large switching field, $V_s\gtrsim1$, the picture is more complicated. The conductance $G(Q_0,q_0)$ is not a periodic function of parameter $q_0$ any more. Instead, some interesting effects related to the breaking and condensation of conductance peaks appear.

The closest peaks to $Q_0=0$ are not only shifted but also reshaped and finally break
at the critical values of parameter $q_0$, Fig.~\ref{figVs1G}. We classify the conductance peaks by their original positions at $q_0=0$ when they located at half-integer $Q_0$. We call peaks located at $Q_0=\pm1/2$ as the first pair of peaks, while peaks sitting at $Q_0=\pm3/2$ -- the second pair of peaks. The first pair of conductance peaks breaks at $q_0=q_0^{c1}$, the second pair breaks at $q_0^{c2}\gtrsim q_0^{c1}$, Fig.~\ref{figVs1G}. Increasing parameter $q_0$ the memory-dependent conductance peaks slowly regroup around $Q_0=0$, some of them even ``collide'' with each other, Fig.~\ref{figVs1G}(d).
We call this system behavior the conductance peak condensation.

We introduce the quantity $\alpha_{\mathrm l}(V)=\partial P/\partial V$.
In the vicinity of zero voltage $\alpha_{\mathrm l} \approx q_0/V_s$. For voltages $V>V_s$ we find
 $\alpha_{\mathrm l} \approx 0$. The relation between the grain charge and the grain potential has
the form $ne=Q_0+\Cs\phi+\int_0^{\phi-V_g}\alpha_{\mathrm l}(u)du$.
The position of conductance peaks can be approximately evaluated using the following relation
\begin{equation}\label{Eq_peaks}
Q_0^n+\int_{0}^{Q_0/C_g}\alpha_{\mathrm l} du=(n+1/2)e.
\end{equation}
In the vicinity of $Q_0=0$ the distance between peaks reduces from $1e$ to $1e/(1+\alpha/C_g)$.
For $Q_0 > V_s C_g$ the interpeak distance becomes $1e$.

Now we focus on the polarization dependence of parameters $Q_0$ and $q_0$
and their correlation with features in the conductance peaks. As follows in
Fig.~\ref{figVs1G}(e)-(h) the polarization also shows the memory effect similar
to the conductance. In Fig.~\ref{figVs1G}(g) one can see a number of hysteresis loops;
their edges exactly correspond to features in the conductance;
the number of the hysteresis loops corresponds to the number of conductance peaks broken and ``condensed''
near $Q_0=0$. At first glance the sharp changes in the polarization
in Fig.~\ref{figVs1G} contradict the chosen parameter $V_s=1$ that makes
the polarization a smooth function, Eq.~\eqref{P-approx}. However,
the features are related to the quantization of the excess charge on the grain of SET
and not to the parameter $V_s$.

In the vicinity of $Q_0^n$ the ground state of the SET is described by Eqs.~(\ref{EqAvQ}), (\ref{EqProb})
and (\ref{EqCharge1}). These equations provide the criterion for hysteresis appearance
\begin{equation}\label{EqTcrit}
T<T^n_{\rm cr}=\frac{E_c^0\alpha_{\mathrm l}}{4(\Cs-\alpha_{\mathrm l}(Q_0^n/C_g))}.
\end{equation}
The hysteresis appears first for peaks in the vicinity of $Q_0=0$ where $\alpha_{\mathrm l}$ is large.
At large $Q_0$ the hysteresis disappears since $\alpha_{\mathrm l}\to 0$.
Similar to step-like polarization behavior the hysteresis disappears with increasing temperature.

In Fig.~\ref{DplotVs10} we investigate the SET with large switching
voltage, $V_s=5$. Qualitatively the conductance behaviour is
similar to the case of $V_s=1$ except the fact that more conductance peaks
are involved in the memory effect for similar values of $q_0$. Figure~\ref{DplotVs10} shows
the evolution of conductance with parameter $q_0=1,3,10$. The density plots for
conductance show a complicated structure, strongly nonperiodic in $Q_0$ unlike the SET
without ferroelectricity.
\begin{figure}[t]
  \centering
  \includegraphics[width=\columnwidth]{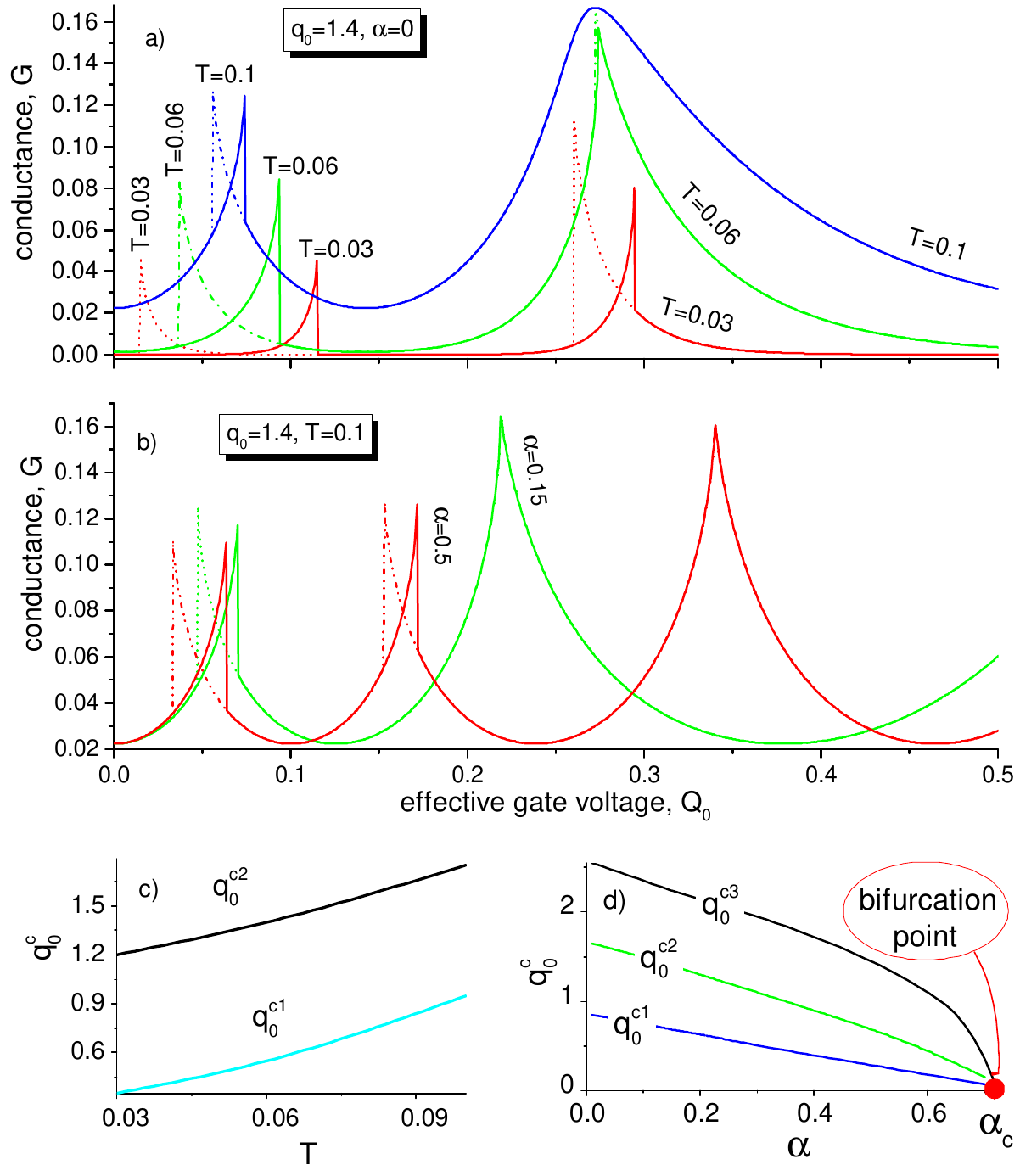}\\
  \caption{(Color online) (a),(b) Memory effect in the conductance for different
  temperatures and parameters $\alpha$. All parameters are the same as in Fig.~\ref{figenergy},
  except $V_s=1$. Branches of $G(Q_0)$ for increasing (decreasing) $Q_0$ are shown
  by solid and discontinuous curves. (c) and (d) Critical $q_0$ where the first and the
  second pairs of conductance peaks undergo the bifurcation and acquire the memory dependence.
  (c) Suppression of $q_c^0$ for finite $\alpha$. (d) Bifurcation point $\alpha_c$ where all $q_c^0$
  merge and go to zero. Conductance $G(Q_0)$ behavior for $\alpha>\alpha_c$ is different from
  its behavior for $\alpha<\alpha_c$. }\label{figGTVs1}
\end{figure}

Analysing numerical data we conclude that conductance peaks condensation
appears for the following gate voltages: $|V_g| - \langle\phi\rangle_{\rm max} < V_s$ or $|Q_0| < C_g (V_s+ \langle\phi\rangle_{\rm max})$, where $\langle\phi\rangle_{\rm max}$ is the maximum grain potential.
At zero temperature, $T=0$, the maximum grain potential
$\langle \phi \rangle_{\rm max}=\frac{|e|}{2C_\Sigma}$. The number of condensed peaks is
approximately equal to the maximum polarization charge that
ferroelectric can induce on the grain, $N_{\rm cond}\approx 2q_0 $.

\subsubsection{Temperature dependence of the memory effect}

The important question is the temperature dependence of the memory effect.
According to Eq.~\eqref{go} the width of the conductance peaks in SET
without ferroelectricity is approximately proportional to the temperature.
Similar effect can be seen with ferroelectric located in the gate
capacitor, see Fig.~\ref{figGTVs1}(a). Moreover, it follows that there is a pronounced temperature
dependence of the critical $q_0$ where the first, second etc... peaks undergo
the bifurcation and acquire the memory dependence. The graphs of $q_0^{c1,2}$ are
shown in the inset of Fig.~\ref{figGTVs1}(c).

\subsubsection{Influence of finite linear term in Eq.~(\ref{P-approx})}
\begin{figure}[t]
  \centering
  \includegraphics[width=\columnwidth]{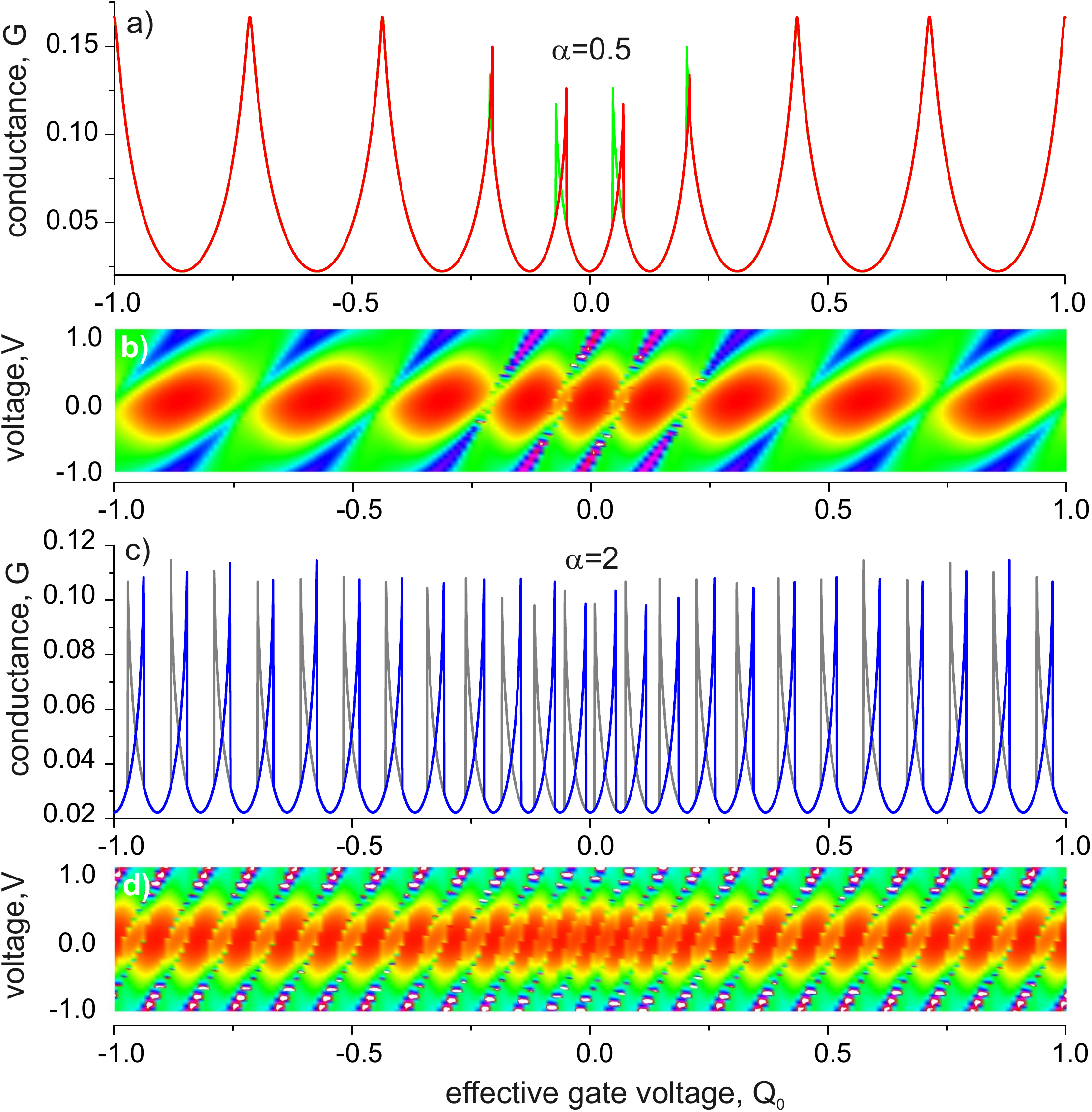}\\
  \caption{(Color online) Conductance $G(Q_0)$ behavior
  for $\alpha>\alpha_c$ is different from its behavior for $\alpha<\alpha_c$.
  Since all $q_c$ are zero above $\alpha_c$ the conductance peaks
  have hysteresis for all $Q_0$. Here $V_s=1$, $T=0.1$, $q_0=1$, $\alpha=0.5<\alpha_c$ in (a) and (b) while $\alpha=2>\alpha_c$ in (c) and (d). Branches of $G(Q_0)$ for increasing (decreasing) $Q_0$ are shown in (a) in red and green curves, while
  in (b) -- in blue and grey curves. }\label{figalphas}
\end{figure}
The presence of the linear term $\alpha$ in Eq.~(\ref{P-approx}) for polarization $P$ is
strongly influence the conductance of SET, Fig.~\ref{figGTVs1}(b). Generally, finite $\alpha$
leads to the renormalization of $\alpha_{\mathrm l}$: $\alpha_{\mathrm l}\to \alpha_{\mathrm l}+\alpha$. Finite $\alpha$ causes the hysteresis behavior of all peaks at small enough temperature.  The presence of small $\alpha$ shifts the
conductance peaks and reduces their amplitudes, while the larger $\alpha$ changes the critical values of parameter $q_0$ where the corresponding conductance peaks undergo the bifurcation and acquire the memory dependence.
Figure~\ref{figGTVs1}(d) shows that the presence of finite $\alpha$ suppresses $q_c^0$.
There is a bifurcation point $\alpha_c$ where all $q_c^0$ merge and reach zero. The behavior of conductance $G(Q_0)$ for $\alpha>\alpha_c$ is very different from its behaviour for
$\alpha<\alpha_c$. This is related to the fact that all $q_c$ are zero above $\alpha_c$ therefore all the conductance peaks for any value of parameter $Q_0$ have the hysteresis, Fig.~\ref{figalphas}.

\begin{figure*}[t]
  \centering
  \includegraphics[width=2\columnwidth]{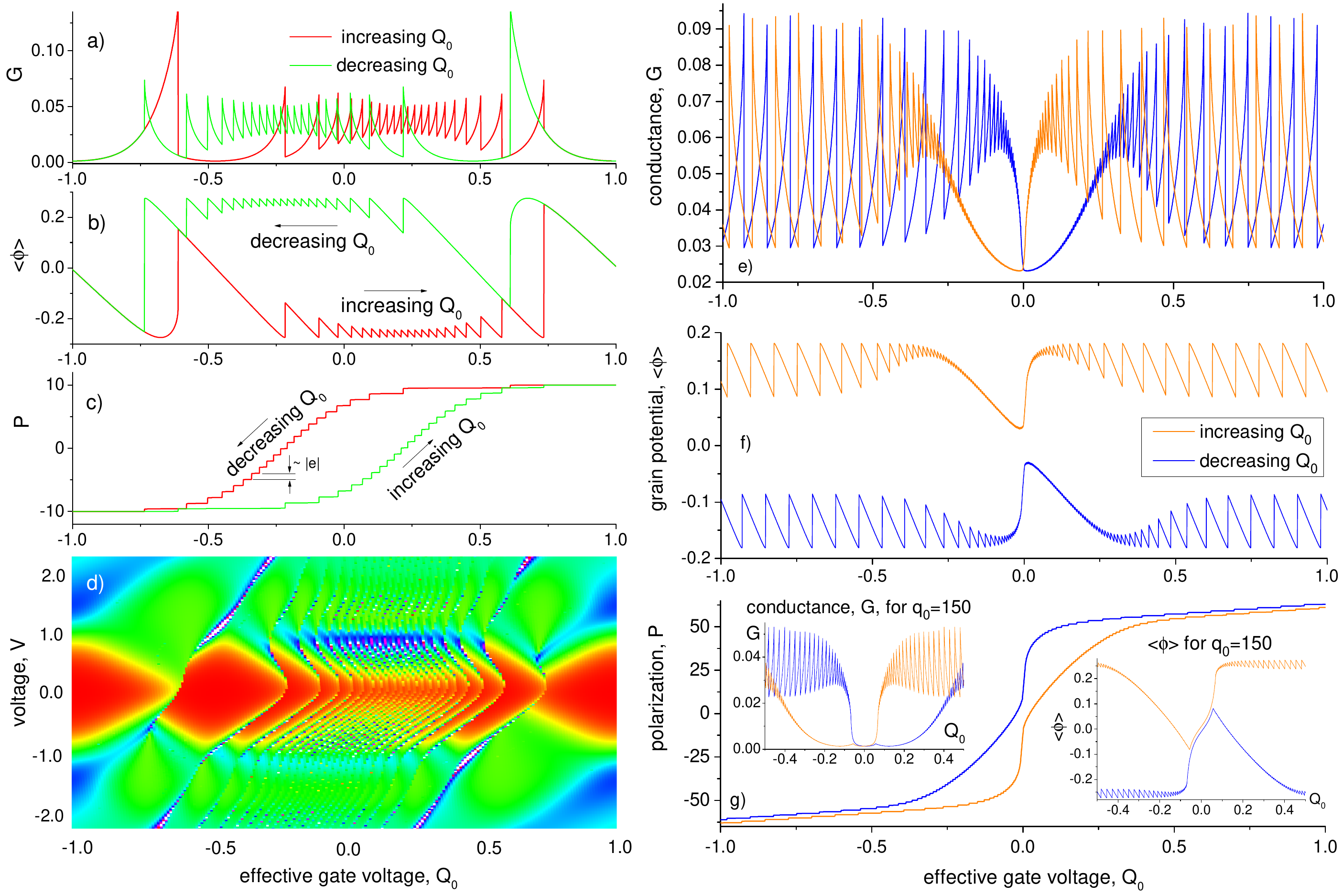}\\
  \caption{(Color online) Coulomb blockade induced memory effect (in the absence of $P(V)$ hysteresis). Here $V_h$=0, $C_1 = 0.1$, $C_2 = 0.05$, $C_g = 0.85$, $q_0 = 10$, $T = 0.06$, $V_s = 0.3$ and  $\alpha= q_0/(30 V_s)$.   There is no hysteresis in $P(\mathcal E)$, see Eq.~\eqref{P-approx}. (a) Conductance,   (b) Average grain potential (Each small step in polarization $P(Q_0)$ in (b) corresponds to the change of the grain charge by the charge quantum), (c) Polarization, and (d) Coulomb-diamond plot (color gradients show the conductance). In (a)-(c) the red curves correspond to the memory-branch with increasing $Q_0$, while the green curve to the decreasing $Q_0$. (d) corresponds to increasing $Q_0$. Plot (e)-(g) shows the conductance, average potential of the grain and polarization for $q_0=50$ and $T=0.1$ ($\alpha$ here is larger $\alpha_c$).  Insets in (g) show $G$ and $\langle\phi\rangle$ for $q_0=150$.}\label{Fighystlarge}
\end{figure*}
The critical $\alpha$ can be found analytically.  We find $\alpha$ corresponding
to an ambiguous solution for the ferroelectric polarization $q_e$ in the limit $Q_0\to\infty$ using
the self-consistency equation $q_g=q_g(V)=q_g(\langle\phi\rangle(Q_0+q_g)+Q_0/C_g)$, Appendix~\ref{Ap1}.
Differentiating in voltage $Q_0$ we find
\begin{equation}\label{eq:selfConsDer}
\frac{d q_g}{d Q_0}=\alpha\left(\frac{d \langle\phi\rangle(Q'_0)}{d Q'_0}\left(1+\frac{d q_g}{d Q_0}\right)+\frac{1}{C_g}\right)
\end{equation}
where $Q'_0=Q_0+q_g$ and $\langle\phi\rangle(Q_0)$ denotes the
average potential of SET without ferroelectric in the gate capacitor.
The ambiguity in the solution  of $q_g(Q_0)$ results in the appearance of singularity in
its derivative. According to Eq.~(\ref{eq:selfConsDer}) the derivative $d q_g/dQ_0$ becomes
singular at some points if
\begin{equation}
\max_{Q_0}\left(\alpha\frac{d \langle\phi\rangle(Q_0)}{d Q_0}\right)>1.
\end{equation}
According to the orthodox theory, the derivative of the average potential approaches its maximum at the
degeneracy points leading to
\begin{equation}
\alpha_c=C_{\Sigma}\left(\frac{E_c}{2T}-1\right)^{-1}.
\end{equation}

Another effect due to the presence of finite $\alpha$ is the
renormalization of the distances between the conductance peaks for $Q_0$ away from zero.
For zero temperature the distances between peaks are reduced by a factor of $\left(1+\frac{\alpha}{C_{g}}\right)/\left(1+\frac{\alpha}{C_{\Sigma}}\right)$.

\subsubsection{Giant hysteresis memory-loop in the absence of $P(V)$ hysteresis ($V_h=0$).}\label{sec:GiantHysteresusLoop}
\begin{figure*}[t]
  \centering
  \includegraphics[width=2\columnwidth]{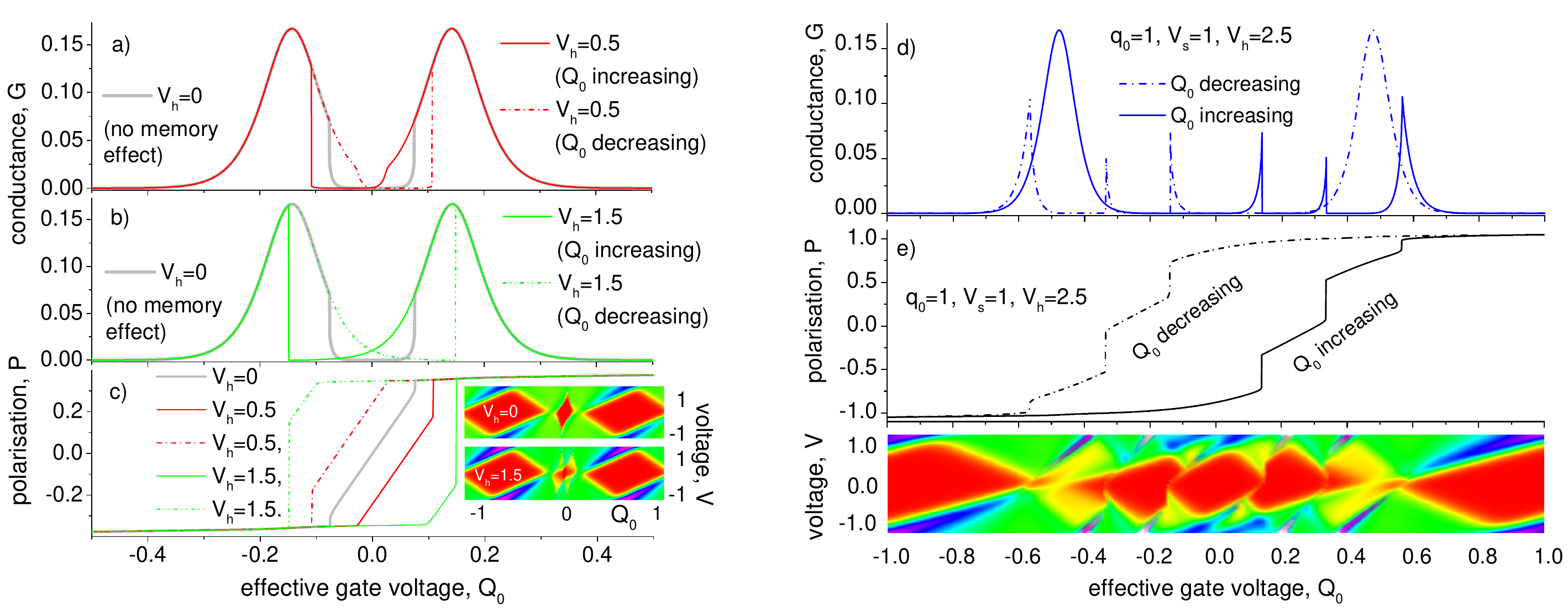}\\
  \caption{(Color online) Conductance and polarization of ferroelectric with finite
  hysteresis. Graphs (a), (b), and (d) compare SETs with $V_h=0$ and $V_h>0$; for $V_h=0$
  the memory effect are absent in $G(Q_0)$. In graphs (a), (b) and (c) parameters
  $T=0.03$, $V_s=0.01$, and $q^{0} = 0.35$ are the same as in
  Fig.~\ref{figenergy}. As follows from (a), (b) and (c) hysteresis in
  polarisation-voltage dependence drives the hysteresis in the conductance. Insets in graph (c):
  conductance diamonds evolution with voltage $V_h$. Graphs (d), (e) and (f)
  show memory effect in $G(Q_0)$ for $V_h=0$.}\label{histVs001}
\end{figure*}
Above we showed that ferroelectricity drives the memory effect. Here we show that it can be ``giant''. This is so if
the effective charge ``induced'' by the ferroelectric at the grain is large enough, for example $q_0=10$, Fig.~\ref{Fighystlarge}. We mention that we still use $V_h=0$ such that the hysteresis in $P(\mathcal E)$ is
absent, Eq.~\eqref{P-approx}. However, $P(Q_0)$ has a large hysteresis. Each small step in $P(Q_0)$ curve in Fig.~\ref{Fighystlarge}(b) corresponds to the change of the grain charge by the charge quantum.
This
large Coulomb blockage hysteresis is further increased for $V_h>0$.
This giant hysteresis memory-loop has potential for applications in memory devices.\par

This effect can be thought of as an extreme case of ``peak condensation''. When enough peaks
are ``condensed'', the conductance behavior acquire the quasi-oscillatory character in the
vicinity of $Q_0=0$. Physically, this corresponds to the situation when a
small change in $Q_0$ causes the number of electrons on the grain to be changed by one
at the expense of the FE polarization. As a result of being a periodic function of the number
of excess electrons on the grain the conductance changes little in this process.
The same is true for the average potential.\par

Below we discuss the SET parameters and conditions to observe
the effect of gigantic hysteresis loop. First, the width of the hysteresis loop is restricted by the gate
capacitance $C_g$ meaning that it is beneficial to have a large $C_g$, $C_g > C_1, C_2$. Second, the switching voltage
$V_s$ should be of the order of 1 for the polarization not to flip to fast with the change of $Q_0$.
Third, the conductance peaks should start merging meaning that there should be enough number
of them (large $q_0$) and their width should be sufficient (not too low temperatures, $T$).

\subsection{Ferroelectric with finite hysteresis loop}

Now we consider a ferroelectric with finite hysteresis in the polarization --- the electric field dependence is given
by Eq.~(\ref{P-approx}). In general the state of FE depends on the whole history of its evolution, however here
we consider only processes with monotonous change of FE polarization where the hysteresis-loop approximation is valid.\par
Figure~\ref{histVs001} shows the conductance and polarization of ferroelectric
with finite width hysteresis loop. Graphs (a), (b), and (d) compare SETs with voltages
$V_h=0$ and $V_h>0$, where for $V_h=0$ there is no memory effect in the conductance $G(Q_0)$.  { As follows the intrinsic hysteresis in the FE polarization-voltage dependence increases the hysteresis in the conductance. \par
This behavior is predictable in comparison to the negligible-hysteresis case. The hysteresis in Eq.~(\ref{P-approx}) is equivalent to the introduction of an additional polarization-induced charge on the grain, where charge is being dependent
on the evolution of $Q_0$.
\begin{equation}
\delta q_g^{\leftrightarrows}=q_0\left(\tanh\left(\frac{V\pm V_h}{V_s}\right)-\tanh\left(\frac{V}{V_s}\right)\right).
\end{equation}
This additional charge vanishes in the limit $Q_0\to\pm\infty$, but even in this limit it causes further
retardation of the FE polarization change with $Q_0$ in the region around zero.
Moreover, the intrinsic FE polarization causes broadening of the interval of $Q_0$ where the state of the FE SET
is not unique. This interval region is estimated as $|Q_0| < C_g (V_s+ \langle\phi\rangle_{\rm max}+V_h)$.
}

\section{Discussion}

Typical experimental parameters of SETs and ferroelectric materials were discussed
in details in our paper~\onlinecite{RefOurPRB}.

\begin{figure*}[t]
  \centering
  \includegraphics[width=2.0\columnwidth]{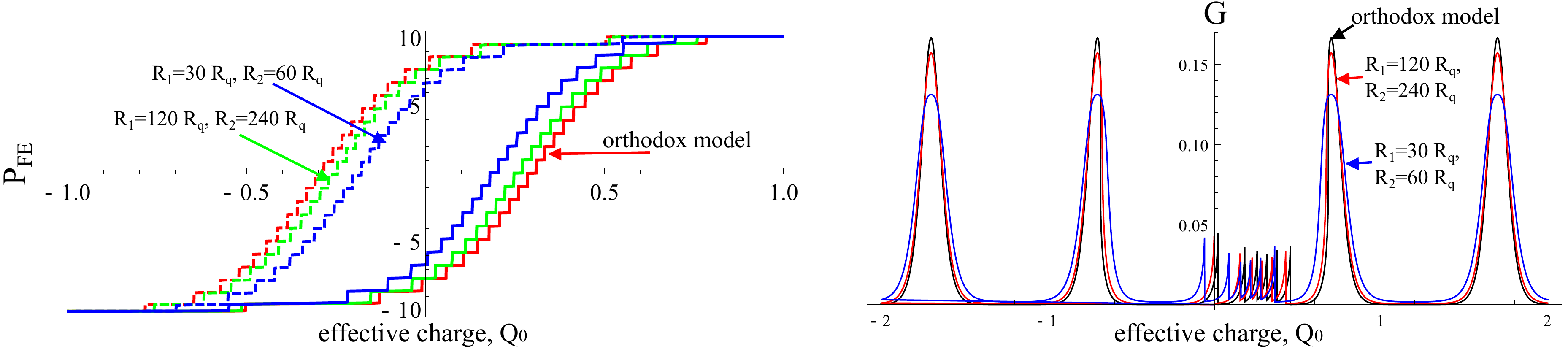} 
  \caption{(Color online) Left figure:  Coulomb-blockade induced hysteresis loop of the gate ferroelectric polarization for different resistances of the tunnel junctions (polarization is expressed in charge units). Right figure: Conductance peaks line shape distortion for different resistances of tunnel junctions (G is normalized to the conductance of
  orthodox model). Parameters: $C_1=0.05$, $C_2=0.1$, $C_g=0.85$, $q_0=10.1$, $V_s=0.3$, $T=0.03$.}
 \label{PFE}
\end{figure*}
\subsection{Influence of cotunneling\label{sec:cotun}}
Above we discussed the SET with ferroelectric gate using the orthodox theory. Now we go beyond this theory
and show how the finite junction conductances influence our results. In particular, we show that the next order
corrections to the SET conductance and the island occupation number do
not qualitatively change the system behavior, but rather introduce some quantitative corrections.\par
The orthodox theory assumes that the junction conductances are small compare to $4\pi^2 e^2/h$ and takes into account only the lowest order processes in the tunneling matrix elements. In this section we use the results of~\cite{PhysRevLett.78.4482}
to calculate the second order corrections to the conductance and the first order correction to the electron
mean occupation number on the island. The latter results in a corrected average island potential $\langle\phi\rangle$ that is used in the self-consistency equation.

In this section we measure the resistance in quantum units $R_q=h/(4 \pi^2 e^2)$.\par

In the theory of SET the second order corrections in tunneling conductance manifest themselves
in reshaping conductance peaks and in fluctuations of the electron number on the island. For
SET with ferroelectric gate the first effect remains, except the fact that the reshaping becomes
more complicated in the vicinity of $Q_0=0$. However, the quantum charge
fluctuations introduce a new effect --- they reduce the polarization hysteresis loop.
This phenomenon stems from the fact that additional fluctuations of the island charge allow
the FE polarization to be switched more easily. However, this effect cannot eliminate the
hysteresis completely because the charge fluctuations are suppressed outside the conductance peaks. The
typical shifts in the polarization hysteresis due to quantum fluctuations are shown in Fig.~\ref{PFE}. \par

\subsection{Memory effect devices}

The presence of hysteresis in transport characteristics of ferroelectric SET has potential applications for memory devices. The memory bit is associated with particular direction of the ferroelectric polarization in the gate capacitor. The memory storage corresponds to the zero gate voltage ($Q_0=0$).
For example, if we measure the conductance at fixed parameter $Q_0 = 0.5$ in Fig.~\ref{histVs001}(d)
it gives the direction of the polarization. This is the reading operation.
The writing information (fixation of a particular polarization direction) can be performed
by applying a large enough gate voltage ($Q_0$).\par
{
The stored information may be strongly influenced by the fabrication-dependent SET parameters such as $q_0$ and $V_s$. For instance, a change in $q_0$ by half of the elementary charge would shift the conductance peaks by half a period, dramatically changing the relationships between conductances and island potentials for opposite FE polarization.\par

For memory application the VDF-TrFE ferroelectric can be used.~\cite{Yamada1981JAP} It has the Curie point above the room temperature. The polarization of this relaxor in the vicinity of room temperature is about $P=3$ mkmC/m$^2$ producing
the charge $q_0\approx0.5$~e for $5nm$ grain size. Moreover, the magnitude of VDF-TrFE
polarization can be tuned varying the VDF concentration. The dielectric permittivity of this relaxor is
rather small in the vicinity of room temperature, $\epsilon^{\mathrm{VDF-TrFE}}\approx5$.
Therefore this ferroelectric can not suppress the Coulomb blockade effect and
the charging energy can be as high as $2000 K$. This allows the memory device
to operate at room temperature. For memory device the FE thickness can be of order $10 nm$.
Assuming that the distance between the grain and the leads is about $1 nm$ we obtain a
good relation between the gate capacitance and the leads capacitors. The switching field of this FE is about $600 kV/cm$
giving $V_\mathrm{h}\approx 2.5$. Using above estimates one can see that the SET with VDF-TrFE
ferroelectric has behavior similar to the one shown in Fig.~\ref{histVs001}. Therefore this
system can be used for memory application.

However, utilizing the giant hysteresis memory loop described
in Sec.~\ref{sec:GiantHysteresusLoop} can provide another configuration,
which is less sensible to the SET parameters. As numerical calculation show (see Fig.~\ref{PFE}), for large
enough  parameter $q_0$, the island potential tends to ``stick'' to its maximum or
minimum value at $Q_0=0$ depending on the way it was set there. Thus, the island
potential may serve as a reliable indicator of the ferroelectric polarization which can be directly measured
using an auxiliary quantum dot.~\cite{ncomms1620}
}

\subsection{Conclusions}

{We investigate electron transport properties of SET with FE insulator placed between the metallic grain and
the gate electrode. The mutual influence of charged grain and the FE polarization leads to drastic changes in
the SET transport. In particular:}
\begin{enumerate}
\item There is an ambiguity in the $I(V)$ characteristics of SET with ferroelectric gate-capacitor
originating from nonlinear mutual influence of electron at the metallic grain and the
FE polarization. It appears even in the absence of hysteresis in the FE polarization.

\item The state of SET is no longer periodic in the gate voltage ($ Q_0 $). In particular,
the  ``condensation'' of conductance peaks appears in the vicinity of $ Q_0 = 0$. The range of $ Q_0 $ where
conductance peaks condensate depends weakly on the ferroelectric properties and other transistor parameters.
For $V_h=0$ the peak condensation appears for gate voltages $|V_g| - \langle\phi\rangle_{\rm max} < V_s$ or $|Q_0| < C_g (V_s+ \langle\phi\rangle_{\rm max})$, where $\langle\phi\rangle_{\rm max}$ is the maximum grain potential. At zero temperature, $T=0$, the maximum potential is $\langle \phi \rangle_{\rm max}=\frac{|e|}{2C_\Sigma}$. The number of the condensed peaks is approximately equal to  the maximum polarization charge that ferroelectric can induce on the grain: $N_{\rm cond}\approx 2q_0 $.

\item The linear part of the ferroelectric polarization $\alpha V$ substantially influences the conductance
behavior as a function of parameter $ Q_0 $. In particular, the finite $\alpha$ leads to the
reduction of conductance peaks and the distances between the peaks in the whole range of
$Q_0$ (not only for small $Q_0$).

\item {A finite hysteresis loop of the FE polarization makes the distinction
between forward and backward change of voltage $V_g$ more pronounced.
A finite voltage $V_h$ is not needed for ambiguity in the SET state to appear.}
\end{enumerate}

\acknowledgments
N.C. acknowledges for the hospitality Laboratoire de Physique Théorique, Toulouse and CNRS where this work was finalized and SIMTECH
Program, New Centure of Superconductivity: Ideas, Materials and Technologies (grant No. 246937). S.~F., A.~K, were supported by Russian Scientific Foundation (Grant No. RNF 14-12-01185), N.C. by RFBR No. 13-02-0057, and I.~B. was supported by NSF under Cooperative Agreement Award No. EEC-1160504, NSF Award No. DMR-1158666, and the NSF PREM Award.

\appendix

\section{Theory of Ferroelectric Single Electron Transistor \label{Ap1}}

\subsection{Self-consistent solution}

In this section we show how the influence of ferroelectric polarization can be included into the theory of SET.
We limit our consideration to the case of sufficiently slow ferroelectric response times $\tau_{P}$ compare to the electron tunneling time $\tau_e=R_{\Sigma}C_{\Sigma}$. The opposite limit is discussed
in~Ref.~\onlinecite{RefOurPRB}. We consider the steady-state solutions.\par

We use the following Hamiltonian to describe the SET
\begin{equation}
H=H_1+H_2+H_I+H_c+H_T
\end{equation}
Here $H_{k}=\sum_{l}\epsilon_{kl}a_{kl}^{\dagger}a_{kl}$, $k=1,2,I$ denotes Hamiltonians for isolated right and left leads and the island respectively, $H_T$ is the tunneling Hamiltonian, and $H_c$ is the Coulomb energy of the form
\begin{equation}
H_c=\frac{1}{2C_{\Sigma}}\left(e\hat{n}-Q_0-(C_1-C_2)\frac{V}{2}\right)^2.
\label{eq:coulombHamiltonian}
\end{equation}
$Q_0=-C_gV_g$ is the effective charge. The notations for voltages and capacitances are similar
to Fig.~\ref{fig_device}.\par

The Coulomb Hamiltonian in Eq.~(\ref{eq:coulombHamiltonian}) treats the dielectric polarizations in the
junctions as classical variables. Thus it can be easily generalized to the case of an
additional FE polarization $P$ by adding the term $\propto \hat{n}P$
\begin{equation}
H'_c=\frac{1}{2C_{\Sigma}}\left(e\hat{n}-Q_0-q_g-(C_1-C_2)\frac{V}{2}\right)^2.
\label{eq:FEcoulombHamiltonian}
\end{equation}
Equation~\eqref{eq:FEcoulombHamiltonian} is valid for "frozen" FE polarization, meaning that it is constant
on the time-scales of tunneling. This limit is justified assuming that the slow polarization is defined by the mean field
and is weakly influenced by the fast field fluctuations of the charge tunneling.
To find a steady state of SET the equilibrium FE polarization $P_{\rm eq}$ must
be taken constant for calculating the electron tunneling rate. This constant polarization yields an
additional constant charge on the island $q_g$
\begin{equation}
q_g=\int_g d\mathbf S_g\cdot\bf{P_{eq}}.
\end{equation}

The Coulomb Hamiltonian in Eq.~(\ref{eq:FEcoulombHamiltonian}) coincide with the usual SET Hamiltonian
if we introduce a new effective gate charge $Q'_0=Q_0+q_g$. Thus for a given FE polarization
the steady state of the FE SET can be calculated using the theory of usual SET. On the other hand, the polarization of the FE by itself is determined by the mean electric field resulting from the microscopical tunneling dynamics. Therefore
the complete steady-state solution for the FE SET is obtained when the equilibrium FE polarization
and the tunneling dynamics are calculated self-consistently.~\cite{RefOurPRB}\par

\subsection{First order theory}

Here we discussed the first order perturbation theory in tunneling.
For the FE induced charge we use the following expression
\begin{equation}
q_g=q_0 \tanh\left(\frac{V\pm V_h}{V_s}\right)+\alpha V.
\label{eq:qgCalc}
\end{equation}
Here $V=\langle\phi\rangle-V_g$ is the voltage across the gate junction. The average island potential $\langle\phi\rangle$ that plays a key role in determining the FE polarization is a linear function of the average island occupation number $\langle n\rangle$
\begin{equation}
\langle\phi\rangle =\frac{1}{C_{\Sigma}}\left(e\langle n\rangle-Q_0-q_g-(C_1-C_2)\frac{V}{2}\right).
\label{eq:phiAvr}
\end{equation}

In the leading order the probability per unit time to change the island occupation
number from $n$ to $n\pm 1$ through the first junction is given by the Fermi golden rule
\begin{equation}
    \Gamma^{(1)}_{n\to n\pm1}=\frac1{e^2 R_1}\cdot \Delta F_1^{n\to n\pm1} N_B\left( \Delta F_1^{n\to n\pm1} \right),
\end{equation}
where $N_B(\omega)=1/[\exp(\omega/T)-1]$ is the Bose-function,~\cite{Chtchelkatchev2013PRB88Universality} $R_1$ is the tunnelling
bare resistance and $\Delta F_1^{n\to n\pm1}$ denotes the free energy change
with $Q'_0$ being the effective charge
\begin{equation}
   \Delta F_1^{n\to n\pm1}=\frac{e^2}{C_{\Sigma}}\left(\frac{1}{2}\pm\left(n-\frac{Q_0'}{e}\right)\pm \frac{(C_2+C_g/2)V}{e}\right),
\end{equation}
The flow rates defined the probabilities $p_n$ to find $n$ excess electrons on the island through the detailed-balance equations
\begin{equation}
p_n\left(\Gamma^{(1)}_{n\to n+1}+\Gamma^{(2)}_{n\to n+1}\right)=p_{n+1}\left(\Gamma^{(1)}_{n+1\to n}+\Gamma^{(2)}_{n+1\to n}\right),
\end{equation}
with boundary conditions $p_{\pm\infty}=0$. In their turn, probabilities $p_n$ are used to calculate the average occupation number and the average potential through Eq.~(\ref{eq:phiAvr}). The last step is to solve the equations~(\ref{eq:qgCalc}) and~(\ref{eq:phiAvr}) together to obtain a self-consistent solution.\par
Knowing the FE induced charge $q_g^{eq}$, the current through the SET can be found
\begin{equation}
I=|e| \sum_n p_n \left(\Gamma^{(1)}_{n\to n+1}-\Gamma^{(1)}_{n\to n-1}\right),
\end{equation}
where all the rates are calculated with $Q'_0=Q_0+q_0^{\rm eq}$ being the effective charge.

\subsection{Higher order corrections}

Here we consider next order corrections to the average occupation number and to the tunneling current.
If corrections for SET without FE are known as a function of the effective charge $Q_0$, these corrections
can be generalized for FE SET. Indeed, to solve the self-consistency
Eqs.~(\ref{eq:qgCalc}) and~(\ref{eq:phiAvr}) we need to know the dependence of $\langle n\rangle$ on $Q_0$
for SET without ferroelectricity. If this dependence is known, $\langle n\rangle(Q_0)=\langle n\rangle^{(0)}(Q_0)+\langle n\rangle^{(1)}(Q_0)+...$ it can be placed in Eq.~(\ref{eq:phiAvr}) with the proper substitution
for $Q_0 \rightarrow Q_0+q_g$.

In the Sec.~\ref{sec:cotun} we used the results of Ref.~\onlinecite{PhysRevLett.78.4482} for low temperatures, where
it was shown that in the first order theory the occupation probabilities of
only two neighboring states ($n$ and $n+1$ excess electrons) significantly deviate
from zero. The correction to the average occupation number is given by
\begin{equation}
\langle n\rangle^{(1)}=\frac{C_{\Sigma}}{e}\frac{\partial}{\partial Q_0}[p^{(0)}_n(\phi_n-\phi_{n-1})+p_{n+1}^{(0)}(\phi_{n+1}-\phi_n)]\,,
\end{equation}
where
\begin{equation}
\phi_n=\sum_{k=1,2} \frac{R_q}{R_k}\Delta F_k^{n\to n+1}\mathrm{Re}\,\mathrm{\Psi}\left[i \frac{\Delta F_k^{n\to n+1}}{2\pi T}\right]\,,
\end{equation}
$p^{(0)}_{n,n+1}$ are the lowest-order state occupation probabilities, $R_q=h/(4\pi^2 e^2)$, $k$ denotes junctions and $\mathrm{\Psi}$ is the digamma function. \par

Following Ref.~\onlinecite{PhysRevLett.78.4482} we introduce the generalized flaw rates to calculate
the corrections to the tunneling current
\begin{gather}
g^{\pm}_{k}(\omega)=\pm\frac{R_q}{R_k}\frac{\omega-\mu_k}{\exp(\pm(\omega-\mu_k)/T)-1},
\\
g^{\pm}(\omega)=\sum_{k=1,2} g^{\pm}_{k}(\omega),\qquad g(\omega)=\sum_{\sigma=\pm} g^{\sigma}(\omega),
\end{gather}
where $\mu_k=\pm e V/2$ --- the chemical potentials of the electrodes. Additionally,
\begin{equation}
\Delta_n=\langle n+1|H'_c|n+1\rangle-\langle n|H'_c|n\rangle.
\end{equation}
We consider the case with $0$ or $1$ excess electrons occupying the island. To calculate the
corrections for arbitrary voltages and FE polarization we shift the parameter $Q'_0$ by $n_{\rm sh}e$ where
\begin{equation}
n_{\rm sh}=\left[\frac{ Q_0 +q_g+ (C_1-C_2)V/2}{e}\right] .
\end{equation}
Therefore, in the following we assume that $0<(Q'_0+ (C_1-C_2)V/2)/e<1$.
The first-order tunneling current can be written as follows
\begin{equation}
I^{(1)}(\Delta_0)=\frac{4\pi^2 e}{h}\frac{g^{+}_2(\Delta_0)g^{-}_1(\Delta_0)-g^{+}_1(\Delta_0)g^{-}_2(\Delta_0)}{g(\Delta_0)}.
\end{equation}
The second-order contribution to the tunneling current is divided into three parts $I^{(2)}(\Delta_0)=\sum^3_{i=1}I^{(2)}_i(\Delta_0)$, where
\begin{multline}
I^{(2)}_1(\Delta_0)=\int d\omega\, I^{(1)}(\omega)g(\omega)\times
\\
\Real[p_0^{(0)}R_{-}(\omega)^2+p_1^{(0)}R_{+}(\omega)^2],
\end{multline}
\begin{gather}
I^{(2)}_2(\Delta_0)=-I^{(1)}(\Delta_0)\int d\omega\, \Real\sum_{\sigma=\pm} g^{\sigma}R_{\sigma}(\omega)^2,
\\
I^{(2)}_3(\Delta_0)=-\frac{\partial I^{(1)}(\Delta_0)}{\partial \Delta_0}\int d\omega\, \Real\sum_{\sigma=\pm} g^{\sigma}R_{\sigma}(\omega),
\end{gather}
where $R_{\pm}(\omega)=1/(\omega-\Delta_0+i0^+)-1/(\omega-\Delta_{\pm1}+i0^+)$ and the poles at $\omega=\Delta$ are regularized as Cauchy's principal values $\Real[1/(x+i0^+)]=P\frac{1}{x}$ and their derivatives $\Real[1/(x+i0^+)^2]=-\frac{d}{dx}P\frac{1}{x}$.

\subsection{Choosing branches}
\begin{figure}[thb]
\includegraphics[width=0.99\columnwidth]{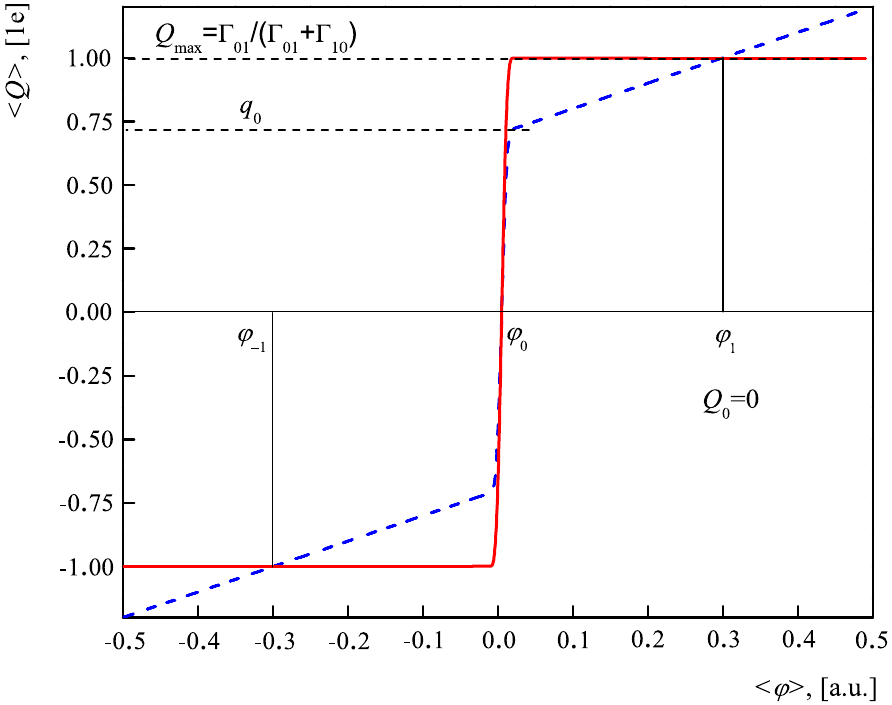}\\
\caption{(Color online) Dependencies Eq.~\ref{EqCharge2} (solid line) and Eq.~\ref{EqCharge1} (dashed line) on $\langle\phi\rangle$. The curves are plotted for the following parameters $q_0=0.55$, $C_g=0.2$, $T=0.03$, $Q_0=0$.}\label{FigCharge1}
\end{figure}
The self-consistent solution of Eqs.~(\ref{eq:qgCalc}) and~(\ref{eq:phiAvr}) may not be unique leading to
the hysteresis behavior discussed in this paper. Thus we encounter the problem of choosing branches.
Here we explain the rules for selecting the solutions.

The self-consistent state is unique for the effective charge being away from the resonance and the
linear part of the FE polarizability $\alpha$ being sufficiently small. There are two situations:
i) $Q_0$ evolved from $-\infty$ to the right and ii) from $\infty$ to the left.
Over the course of $Q_0$ evolution at some values of parameter $Q_0$ the solution can
become discontinuous. In this case we assume that the polarization ``jumps''
to the nearest available position. There is no ambiguity in choosing the new state,
because the polarization always evolves monotonically and takes its maximum
or minimum value depending on the direction of evolution.

\subsection{Ground state of FE SET with step-like polarization}\label{sec:App_Hyst1}

For calculating the ground state of the system all states are important.
However, the hopping probability decreases exponentially with increasing the free energy of a state.
Therefore we consider only few states which are the closest to the considered point.
The number of these states depends on the parameter $q_0$ and the considered point.
For $q_0<0.5e$ only two states $n$ and $n+1$ are important for calculating $\langle\phi\rangle$ for any $n$.
Increasing parameter $q_0$ leads to moving of energy intersection points to $Q_0=0$.
For $q_0 > 0.5$ the situation is more complicated.
In the vicinity of $Q_0=0$ a lot of intersections of different energy branches occur
and one has to take into account many different states. However, for large values of $Q_0$ only two states of
energy intersection points are important.

We consider the case when two states are enough to describe
the system state and concentrate on the vicinity of the point $Q_0=e(n+1/2)-q_0$ with $Q_0>0$ and $q_0<0.5e$.
The average charge is give by the expression
\begin{equation}\label{EqAvQ}
\langle ne\rangle=\langle Q\rangle=e(nW_n+(n+1)W_{n+1}),
\end{equation}
where $W_{n,n+1}$ are the probabilities for the system to be in the state with $n$ and $n+1$ electrons on the grain.
If only two states are present we find
\begin{align}\label{EqProb}
W_n&=\frac{\Gamma_{n,n+1}}{\Gamma_{n,n+1}+\Gamma_{n+1,n}},
\\\notag
W_{n+1}&=\frac{\Gamma_{n+1,n}}{\Gamma_{n,n+1}+\Gamma_{n+1,n}},
\\\notag
\Gamma_{n,n+1}&=\frac{E_c^0(n+1/2-Q_0/e-S\,P(\langle \phi \rangle-V_g)/e)}{e^{\beta E_c^0(n+1/2-Q_0/e-S\,P(\langle \phi \rangle-V_g)/e)}-1},
\\\notag
\Gamma_{n+1,n}&=\frac{-E_c^0(n+1/2-Q_0/e-S\,P(\langle \phi \rangle-V_g)/e)}{e^{-\beta E_c^0(n+1/2-Q_0/e-S\,P(\langle \phi \rangle-V_g)/e)}-1},
\end{align}
where $\beta=T^{-1}$ and the parameter $S$ is approximately the area of the grain in contact with the ferroelectric. Averaging Eq.~(\ref{EqCharge}) we find the following relation
\begin{equation}\label{EqCharge1}
\langle Q\rangle=\Cs\langle\phi\rangle+Q_0+SP(\langle\phi\rangle-V_g).
\end{equation}
Equations~(\ref{EqAvQ}) and (\ref{EqCharge1}) have always a single solution. Therefore there is no hysteresis in the system for $q_0<0.5e$ and arbitrary $Q_0$.
The conductivity of SET has maximum at $W_n=W_{n+1}$. These maximums are located at point $Q_0=e(n+1/2)-q_0$. The average potential at these points is $\langle\phi\rangle=0$.

For $q_0>0.5e$ and $Q_0>0.5e$ there is no hysteresis. In the vicinity
of $Q_0=0$ the system shows the hysteresis and can not be described by
two states. For $q_0\sim0.5e$, three states $n=0,\pm1$ need to be taken into
account in the vicinity of $Q_0=0$, see Fig.~\ref{figenergy}.
For $q_0\sim1.5e$, five states $n=0,\pm1,\pm2$, etc.

For $q_0\sim0.5e$ the system probabilities with $n=0,\pm1$ have the form
\begin{equation}\label{EqProb1}
\left\{\begin{array}{l}
W_{-1}=\frac{\Gamma_{0,-1}}{\Gamma_{-1,0}}W_0,\\
W_{1}=\frac{\Gamma_{0,1}}{\Gamma_{1,0}}W_0,\\
W_0=\left(1+\frac{\Gamma_{0,-1}}{\Gamma_{-1,0}}+\frac{\Gamma_{0,1}}{\Gamma_{1,0}}\right)^{-1}.
\end{array}\right.
\end{equation}
The average charge $\langle Q\rangle$ is give by the following expression
\begin{equation}\label{EqCharge2}
\langle Q\rangle=e(W_1-W_{-1}).
\end{equation}
Solving Eqs.~(\ref{EqCharge2}) and (\ref{EqCharge1}) we find the
average potential $\langle \phi \rangle$. Figure~\ref{FigCharge1} shows the dependence Eq.~\ref{EqCharge2} (solid line) and Eq.~\ref{EqCharge1} (dashed line) on $\langle\phi\rangle$ for $Q_0=0$.
In this figure the curves have three intersections
at points $\phi_{-1,0,1}$ meaning that the system has hysteresis. The asymptotic behavior of
Eq.~(\ref{EqCharge2}) for $Q_0=0$ is the following
\begin{equation}\label{EqCarge3}
Q_{\mathrm{max}}(T)=\pm\frac{e\Gamma_{0,1}}{\Gamma_{0,1}+\Gamma_{1,0}}.
\end{equation}

The value $Q_{\mathrm{max}}$ depends on temperature and decreases with increasing $T$. At temperature $ T = T_{h}$
it becomes smaller than $q_0$. For temperature $T>T_h$ the hysteresis disappears. However,
at high enough temperatures all states needs to be taken into account and the description with three states only
($n=0,\pm1$) can be incorrect.

\bibliography{our_bib}

\begin{thebibliography}{27}
\bibitem [{foo()}]{footnote}%
  \BibitemOpen
  \href@noop {} {\bibinfo  {journal} {The jumps in the conductance and
  polarization in Fig.~\ref{figenergy} for $q_0<0.5$ are also related to the
  memory effect instability but with rather thin hysteresis loop}\
  }\BibitemShut {NoStop}%
\end{thebibliography}

\end{document}